\documentclass[fleqn,10pt]{wlscirep}
\usepackage[utf8]{inputenc}
\usepackage[T1]{fontenc}
\usepackage{graphicx}	% Including figure files
\usepackage{amsmath}	% Advanced maths commands
\usepackage{amssymb}	% Extra maths symbols

\newcommand{\kms}{\mbox{$\mathrm{km\,s^{-1}}$}}

\newcommand{\msun}{\mbox{$\mathrm{M_{\odot}}$}}
\newcommand{\rsun}{\mbox{$\mathrm{R_{\odot}}$}}

\title{A pulsating white dwarf in an eclipsing binary}

\author[1,*]{Steven G. Parsons}
\author[1]{Alexander J. Brown}
\author[1]{Stuart P. Littlefair}
\author[1,2]{Vikram S. Dhillon}
\author[3]{Thomas R. Marsh}
\author[4]{J. J. Hermes}
\author[5]{Alina G. Istrate}
\author[6]{Elm\'e Breedt}
\author[1]{Martin J. Dyer}
\author[3]{Matthew J. Green}
\author[1]{David I. Sahman}
\affil[1]{Department of Physics \& Astronomy, University of Sheffield, Sheffield S3 7RH, UK}
\affil[2]{Instituto de Astrof\'{i}sica de Canarias, Via Lactea s/n, La Laguna, E-38205 Tenerife, Spain}
\affil[3]{Department of Physics, Gibbet Hill Road, University of Warwick, Coventry, CV4 7AL, UK}
\affil[4]{Department of Astronomy, Boston University, 725 Commonwealth Ave. Boston, MA 02215, USA}
\affil[5]{Department of Astrophysics, Radboud University Nijmegen, P.O. Box 9010, NL-6500 GL Nijmegen, the Netherlands}
\affil[6]{Institute of Astronomy, University of Cambridge, Madingley Road, Cambridge CB3 0HA, UK}
\affil[*]{s.g.parsons@sheffield.ac.uk}

\begin{abstract}
White dwarfs are the burnt out cores of Sun-like stars and are the final fate of 97\% of all stars in our Galaxy. The internal structure and composition of white dwarfs are hidden by their high gravities, which causes all elements, apart from the lightest ones, to settle out of their atmospheres. The most direct method to probe the inner structure of stars and white dwarfs in detail is via asteroseismology. Here we present the first known pulsating white dwarf in an eclipsing binary system, enabling us to place extremely precise constraints on the mass and radius of the white dwarf from the light curve, independent of the pulsations. This $0.325\,\msun$ white dwarf --- one member of SDSS\,J115219.99+024814.4 --- will serve as a powerful benchmark to constrain empirically the core composition of low-mass stellar remnants and investigate the effects of close binary evolution on the internal structure of white dwarfs.

\end{abstract}
\begin{document}

\flushbottom
\maketitle

\thispagestyle{empty}

\section*{Introduction}

There are thought to be of order 100-300 million close pairs of white dwarfs in the Galaxy\cite{Maxted99,Nelemans01}. Among these are likely the progenitors of Type Ia supernovae\cite{Tutukov79,Webbink84}, as well as the progenitors of AM CVn binaries\cite{Paczynski67}, R CrB stars\cite{Iben96} and single hot subdwarf stars\cite{Han03}. Moreover, double white dwarf binaries are expected to be the dominant source of gravitational waves for space-based detectors such as the Laser Interferometer Space Antenna (LISA)\cite{Nelemans09,Marsh11}. The core composition of white dwarfs in close binaries can have a profound impact on the outcome of their mergers and any resultant supernovae (for example, white dwarfs with oxygen-neon cores are not expected to explode at the Chandrasakar limit but instead will collapse to neutron stars\cite{Nomoto91}), so there is a need to determine interior structures observationally. Some progress has been made recently through precise mass and radius measurements of white dwarfs in eclipsing binaries\cite{Parsons17}, but ultimately it is asteroseismology that offers the best probe of white dwarf interiors\cite{Fontaine08,Winget08,Althaus10}, capable for example of constraining the carbon/oxygen ratio in white dwarf cores\cite{Charpinet19} and of measuring helium and hydrogen layer masses\cite{Clemens17}. Asteroseismology is also the most definitive way to test for the existence of ``hybrid''-core white dwarfs (carbon-oxygen cores but up to 25 per cent helium by mass), where the distinction from standard carbon-oxygen core white dwarfs is rather subtle\cite{Zenati19}.

Helium-core, low-mass carbon-oxygen and hybrid core white dwarfs are all expected to be particularly prevalent amongst the low mass white dwarf products of binary evolution\cite{Zenati19}, making pulsating white dwarfs in binaries with masses in the range $0.3$ to $0.5\,\msun$ of particular interest. Amongst the $\sim 300$ known pulsating white dwarfs\cite{Corsico19}, very few are in binaries. In fact, apart from a handful of unusual and extremely low mass ($< 0.2\,\msun$) white dwarf pulsators \cite{Corsico14,Bell17}, there is only one known pulsating white dwarf in a detached binary\cite{Pyrzas15}, and at $0.6\,\msun$ it is almost certainly a standard carbon-oxygen core white dwarf\cite{Hermes15}.

In this paper we report the discovery of pulsations from the cooler component of the eclipsing double white dwarf binary SDSS\,J115219.99+024814.4 (hereafter SDSS\,J1152+0248), making it the first known pulsating white dwarf in an eclipsing binary. SDSS\,J1152+0248 is a 2.4 hour binary consisting of two low-mass white dwarfs discovered using data from the {\it Kepler} K2 mission\cite{Hallakoun16}. Preliminary estimates of the parameters of the white dwarfs implied that the cooler component may be close to the ZZ\,Ceti instability strip\cite{Hallakoun16}, but these estimates were highly uncertain due to the fact that only one white dwarf (the hotter) could be seen spectroscopically (i.e. the system appeared single-lined). The higher-quality data presented in this paper unambiguously reveal the system as double-lined, allowing us to measure both masses directly. Moreover, our high time-resolution light curves reveal pulsations from the cooler white dwarf with at least three significant periods (1314, 1069 and 583 seconds) and allow us to constrain the radii of both white dwarfs to high precision (by modelling the eclipse profiles), independent of any theoretical models. Both white dwarfs may have either helium or hybrid helium-carbon-oxygen cores.

\section*{Results}

Following the discovery of SDSS\,J1152+0248 we obtained time-series spectroscopy of the system with the X-shooter echelle spectrograph\cite{Vernet11} on the 8.2\,m Very Large Telescope, with the aim of detecting the cooler white dwarf and measuring the radial velocity curves of both components. In addition to this, we obtained 108 minutes of high-speed photometry of the binary using HiPERCAM\cite{Dhillon18} on the 10.4\,m Gran Telescopio Canarias, which revealed not only the primary and secondary eclipses but also pulsations from the cooler component. Combining both these data sets allows us to measure the stellar and binary parameters of SDSS\,J1152+0248, which are provided in Table\,\ref{tab:params}. 

\begin{table}[ht]
\centering
\begin{tabular}{@{}lccccc@{}}
\hline
Parameter & units & value & uncertainty  \\
\hline
Orbital period & days      &  0.099\,865\,265\,42 & $\pm$0.000\,000\,000\,10 \\
Centre of WD$_1$ eclipse & MJD(BTDB) & 57460.651\,021\,8 & $\pm$0.000\,001\,0 \\
$K_1$ & $\mathrm{km\,s^{-1}}$ & 190.6 & $\pm$1.5 \\
$K_2$ & $\mathrm{km\,s^{-1}}$ & 212.3 & $\pm$10.5 \\
$\gamma_1$ & $\mathrm{km\,s^{-1}}$ & 60.2 & $\pm$1.0 \\
$\gamma_2$ & $\mathrm{km\,s^{-1}}$ & 55.7 & $\pm$7.2 \\
{\it Gaia} parallax\cite{Gaia18} & mas & 1.40 & $\pm$0.26 \\
Bayesian distance\cite{Bailer18} & pc & 706 & -119 +175 \\
Right ascension & deg (J2000) & 178.08337 & \\
Declination & deg (J2000) & 2.80399 & \\
SDSS $g$ & mag & 18.32 & $\pm$0.01 \\
\hline
Binary inclination & degrees & 89.44 & -0.03 +0.04 \\
Binary separation & \rsun & 0.799 & $\pm$0.010 \\
Mass ratio & M$_1$ / M$_2$ & 1.11 & $\pm$0.025 \\
$M_1$ & \msun & 0.362 & $\pm$0.014 \\
$M_2$ & \msun & 0.325 & $\pm$0.013 \\
$R_1$ & \rsun & 0.0212 & $\pm$0.0003 \\
$R_2$ & \rsun & 0.0191 & $\pm$0.0004 \\
$\log{g}_1$ & dex & 7.344 & $\pm$0.014 \\
$\log{g}_2$ & dex & 7.386 & $\pm$0.012 \\
T$_\mathrm{eff,1}$ (eclipse) & K & 20800 & -960 +1060 \\
T$_\mathrm{eff,1}$ (SED) & K & 21200 & -1100 +1200 \\
T$_\mathrm{eff,2}$ (eclipse) & K & 10400 & -340 +400 \\
T$_\mathrm{eff,2}$ (SED) & K & 11100 & -770 +950 \\
$E(B-V)$ & mag & 0.07 & $\pm$0.02 \\
\hline
\end{tabular}
\caption{\label{tab:params} Stellar and binary parameters for SDSS\,J1152+0248. The subscript 1 refers to the brighter, hotter, more massive white dwarf, while 2 refers to the fainter, cooler white dwarf, which is the pulsator. We list the effective temperatures for each white dwarf as determined from both the eclipse light curves and the spectral energy distribution fit. The ephemeris values were derived from a combination of our light curve fit and previous measurements\cite{Hallakoun16}. Quoted uncertainties represent the one-sigma confidence intervals.}
\end{table}

\subsection*{The radial velocity semi-amplitudes of both white dwarfs}

\begin{figure}[ht]
\centering
\includegraphics[width=0.9\linewidth]{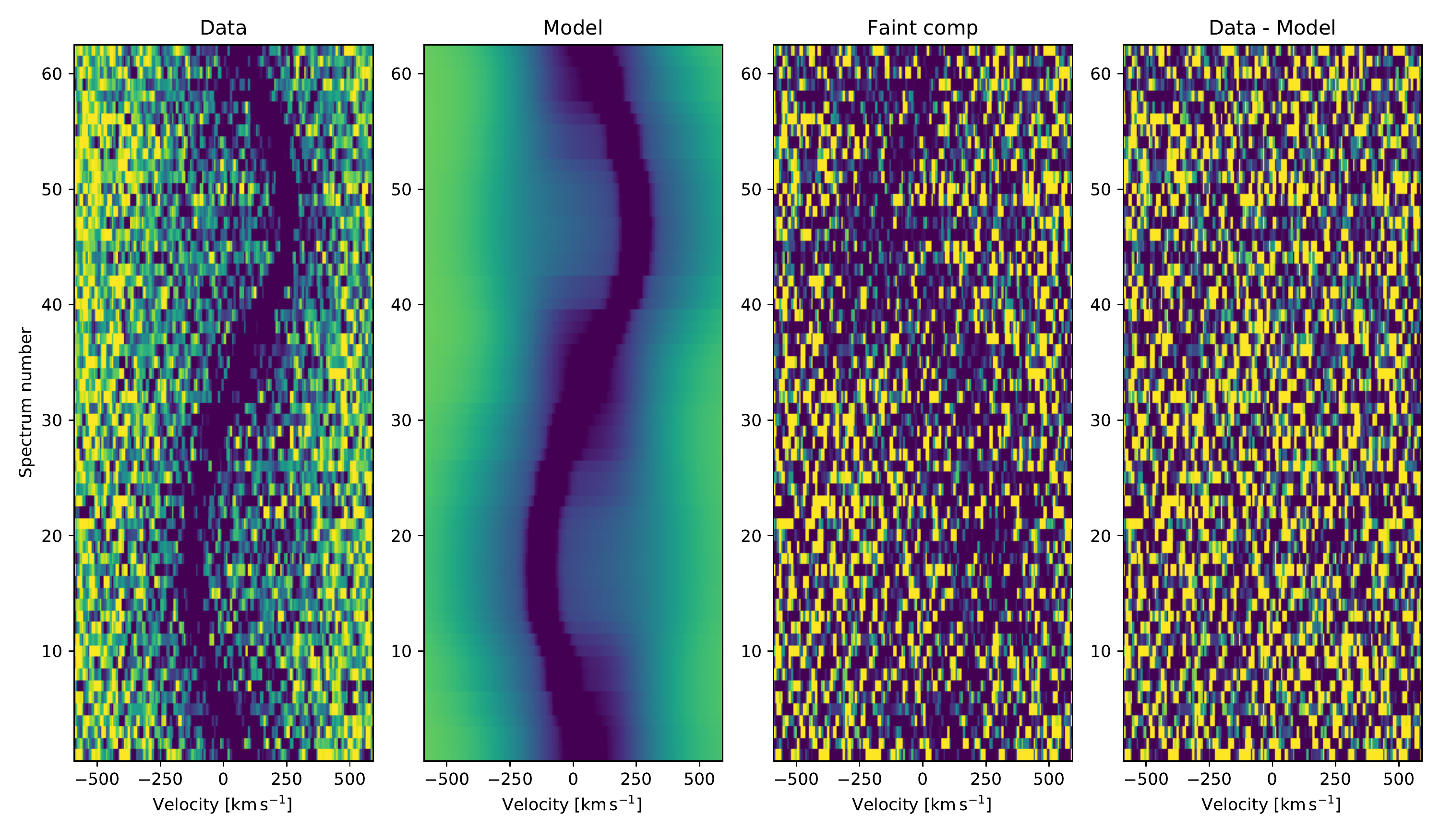}
\caption{Trailed spectrogram of the H$\alpha$ line of SDSS\,J1152+0248. The left-most panel shows the X-shooter data, where the spectra have been ordered in ascending orbital phase (starting and ending at phase 0). The strong absorption line from the hot white dwarf is clearly evident. The next panel shows our best fit model, which includes components from both white dwarfs. The third panel shows the X-shooter data after subtracting the best fit model with only the brighter component included (i.e. it shows only the data for the fainter component). The right-most panel shows the residuals to the fit after including the fainter component as well. The colour scale in the two right-hand plots has been stretched using a $\sinh$ stretch to emphasise the lower fluxes of the absorption line from the fainter white dwarf.}
\label{fig:trail_fit}
\end{figure}

The radial velocity semi-amplitudes for both white dwarfs were determined by fitting the H$\alpha$ absorption line, specifically the narrow non-LTE core of the line. On first inspection the line from the fainter white dwarf is not obvious in the data (the left-hand panel of Figure~\ref{fig:trail_fit}), given that it only contributes $\sim$20\% of the light at this wavelength. However, when a good model for the brighter component is subtracted then the absorption from the fainter white dwarf is seen moving in anti-phase (the third panel of Figure~\ref{fig:trail_fit}). We therefore fitted the X-shooter data with components from both white dwarfs (see the Methods section for a full description of the procedure). We calculated the orbital phase of each spectrum using the ephemeris determined from our light curve fitting (see the next section) and then fitted all spectra simultaneously, shifting the absorption components according to the orbital phase. This approach is useful for determining the velocity of the fainter component, which is often difficult to fit in an individual spectrum. The best fit model is shown in the second panel of Figure~\ref{fig:trail_fit} and the residual of the fit in the right-most panel. The radial velocity values are listed in Table~\ref{tab:params} and indicate a binary mass ratio close to unity.

\subsection*{The masses, radii and effective temperatures of both white dwarfs}

\begin{figure}[ht]
\centering
\includegraphics[width=0.95\linewidth]{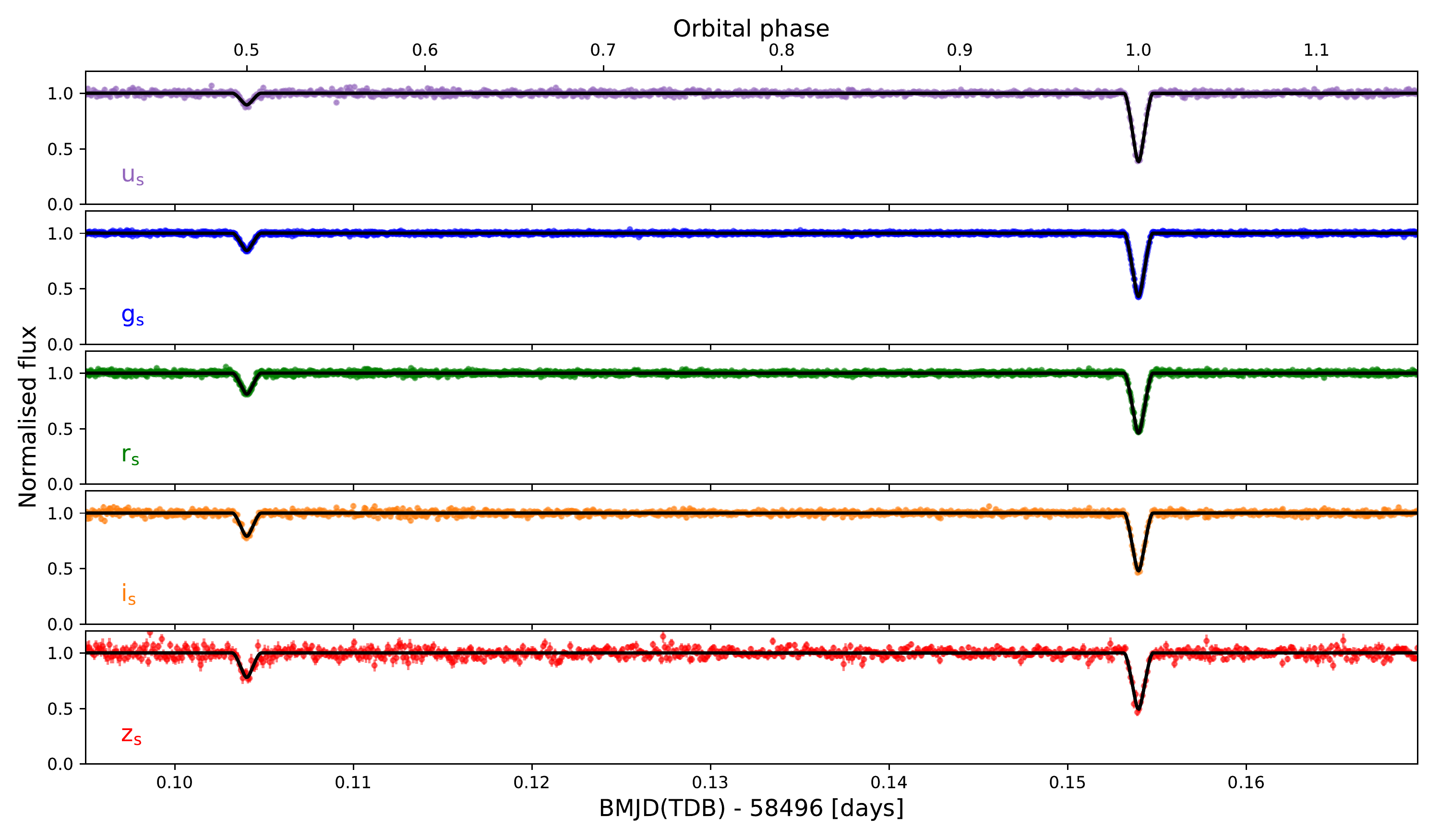}
\caption{HiPERCAM high-speed light curves of SDSS\,J1152+0248 with model fits overplotted. The eclipse of the fainter (cooler) white dwarf occurs at phase 0.5, while the eclipse of the brighter (hotter) white dwarf occurs at phase 1.0. The pulsations from the fainter white dwarf are not visible at this scale.} 
\label{fig:lcurve_fit}
\end{figure}

The high-speed HiPERCAM light curves of SDSS\,J1152+0248 (Figure~\ref{fig:lcurve_fit}) reveal both the primary eclipse (the eclipse of the hotter white dwarf at phase 1.0) and the secondary eclipse (the eclipse of the fainter white dwarf at phase 0.5). Combined with the radial-velocity data from the X-shooter spectroscopy, the light curves allow us to measure the masses and radii of both white dwarfs directly, virtually independent of theoretical models. Moreover, the multi-band light curve data (covering the entire optical wavelength range) also constrains the effective temperatures of the two white dwarfs via the differences in eclipse depths in the different bands.

We used {\sc lcurve}\cite{Copperwheat10} to fit the HiPERCAM data of SDSS\,J1152+0248, which is a code designed to fit the light curves of white dwarf binaries (see the methods section for full details of the light curve fitting procedure). Note that although the pulsations from the cooler white dwarf are not visible in Figure~\ref{fig:lcurve_fit} (due to the plot scale), they are sufficiently large (see the next section) that they needed to be taken into account when fitting the light curves or they would have resulted in systematic errors on the final fit parameters. Therefore, we modelled the pulsations using Gaussian Processes\cite{Rasmussen06}, which are ideal for modelling periodic signals that are not exactly sinusoidal (see the methods section for more details). The derived parameters from the light curve fit and their errors are listed in Table~\ref{tab:params}. The best fit models are shown in Figure~\ref{fig:lcurve_fit}. The measured mass of the cooler white dwarf is lower than in the discovery paper\cite{Hallakoun16} due to an overestimation of the temperatures of the two white dwarfs in that paper. More reliable temperature estimates were only possible due to the combination of our direct detection of the lines from the cooler white dwarf and our multi-band photometric data.

\begin{figure}[ht]
\centering
\includegraphics[width=0.9\linewidth]{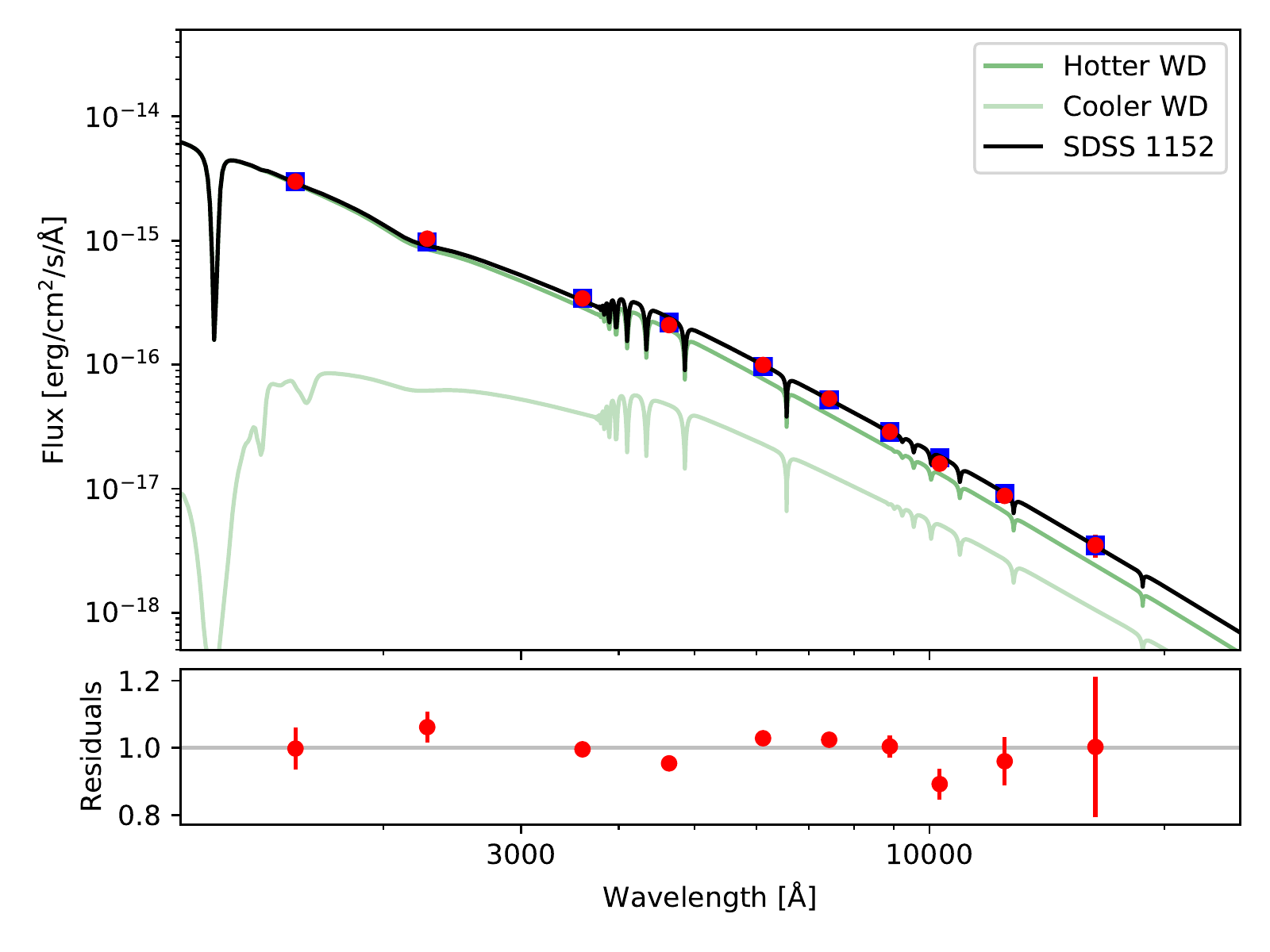}
\caption{Spectral energy distribution of SDSS\,J1152+0248 (red points are GALEX, SDSS and UKIDSS measurements) with best fit model spectrum (black line and blue squares). The model spectrum is a combination of two white dwarf models (individual components shown in green) that have been scaled to match the contribution of each white dwarf in the optical based on the fits to the HiPERCAM eclipse light curves. The lower panel shows the residuals to the fit.} 
\label{fig:sed_fit}
\end{figure}

In addition to the light curve fit we also constrained the temperatures of the two white dwarfs by fitting the spectral energy distribution (SED) of SDSS\,J1152+0248. We retrieved archival photometry from GALEX (FUV and NUV bands), SDSS ($ugriz$ bands) and UKIDSS ($YJH$ bands) and fitted these data with a combination of two DA white dwarf spectra\cite{Koester10} with the same brightness ratios as measured from the HiPERCAM light curves (see the methods section for details of this fit). The result is shown in Figure~\ref{fig:sed_fit} and gives temperatures consistent with the fit to the multi-band eclipse light curves.

\subsection*{The pulsations of the cool white dwarf}

\begin{figure}[ht]
\centering
\includegraphics[width=0.57\linewidth]{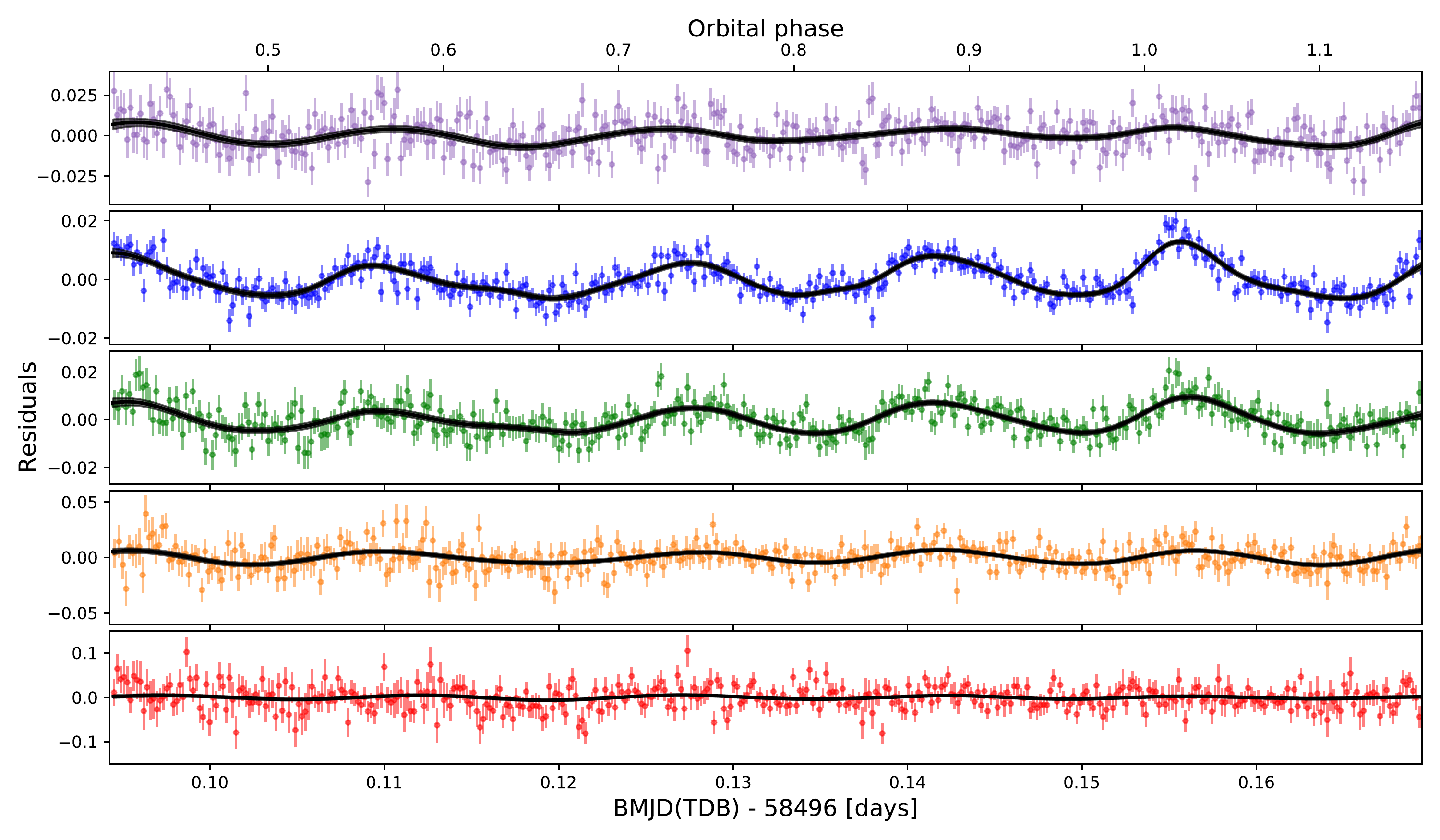}
\includegraphics[width=0.42\linewidth]{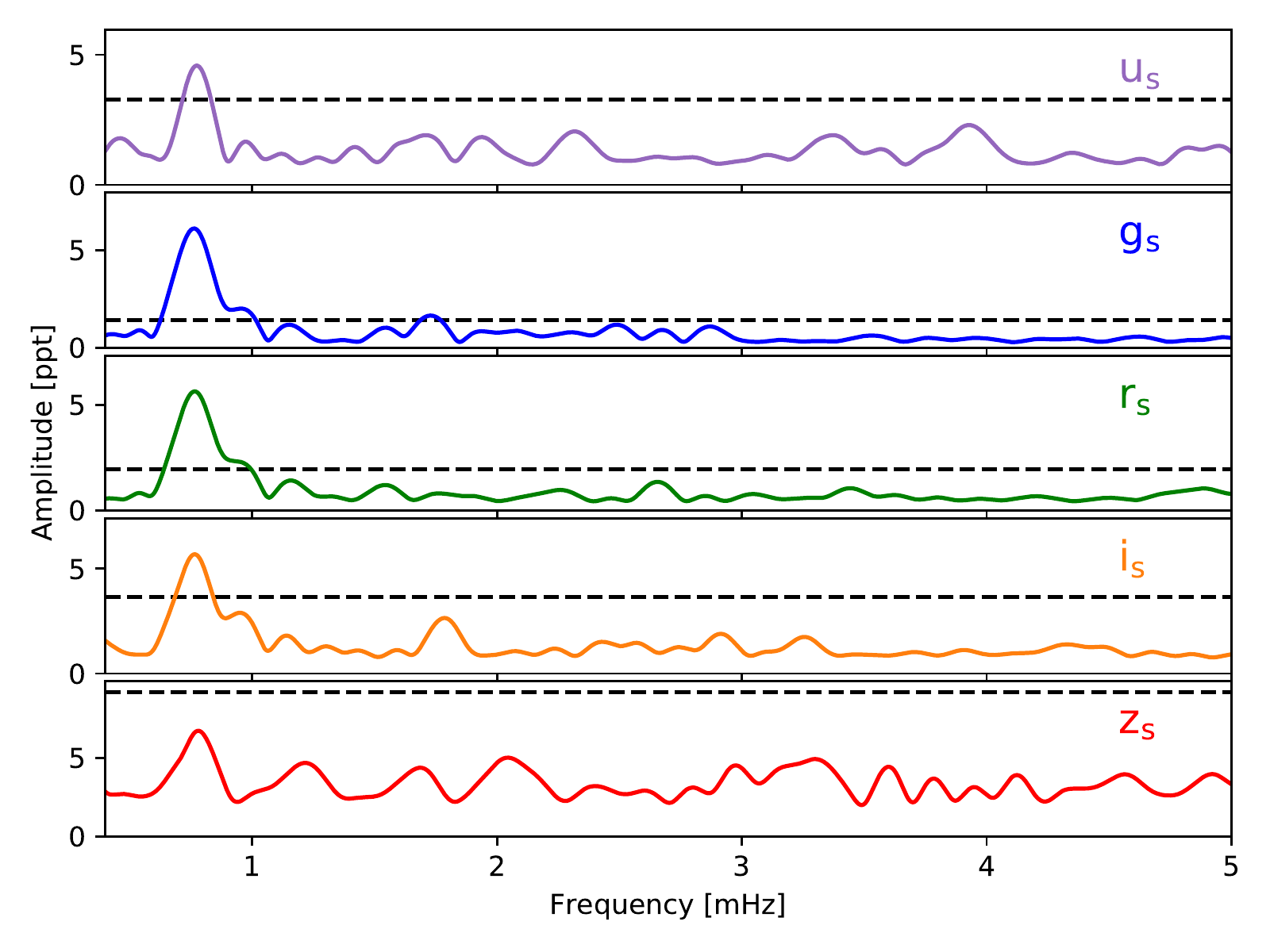}
\caption{The pulsations of the cool white dwarf in SDSS\,J1152+0248. {\it Left:} HiPERCAM light curve residuals after subtraction of the binary model in $u_s$\,$g_s$\,$r_s$\,$i_s$\,$z_s$, respectively. The data have been binned in time by a factor of 5 for clarity. The pulsations are most evident in the $g_s$ band data. Overplotted is the best fit Gaussian Process model that was used to account for the pulsations when fitting the light curve. {\it Right:} Periodograms of the residual light curves showing a clear peak at a frequency of 0.76mHz (period of 1314 seconds), with additional peaks at 0.94mHz (period of 1059 seconds, significant in the $g_s$ and $r_s$ bands) and at 1.71mHz (period of 583 seconds, significant only in the $g_s$ band). The horizontal dashed lines show the 1\% false-alarm probabilities in each band (see the methods section for details of how these were calculated).} 
\label{fig:pulsations}
\end{figure}

The HiPERCAM light curves show small scale periodic variations most easily seen in the $g_s$ band (see the left-hand panel of Figure~\ref{fig:pulsations}). A periodogram\cite{Scargle82} of the data (with the eclipses removed using the binary light curve solution) reveals a strong peak at a frequency of 0.76\,mHz, corresponding to a period of $1314.0\pm5.9$ seconds (the right-hand panel of Figure~\ref{fig:pulsations}), which is seen in all bands, although the detection is only marginal in the $z_s$ band. There are also other peaks at frequencies of 0.96\,mHz (period of $1069.0\pm12.5$ seconds, which is detected in the $g_s$ band and marginally in the $r_s$ band) and at 1.71\,mHz (period of $583.4\pm4.3$ seconds, detected only in the $g_s$ band). These pulsations are in line for those expected for non-radial $g$-mode pulsations of a $\sim$0.3\,\msun\ white dwarf\cite{Corsico14}. The short baseline of the HiPERCAM observations makes a more detailed asteroseismic analysis impossible at this time. 

With a mass of $0.325\pm0.013$\,\msun the pulsating white dwarf in SDSS\,J1152+0248 sits in a poorly sampled region of the ZZ\,Ceti instability strip between standard carbon-oxygen core ZZ\,Ceti white dwarfs and the extremely low-mass helium-core pulsators\cite{Kilic18}.

\section*{Discussion}

\begin{figure}[ht]
\centering
\includegraphics[width=0.47\linewidth]{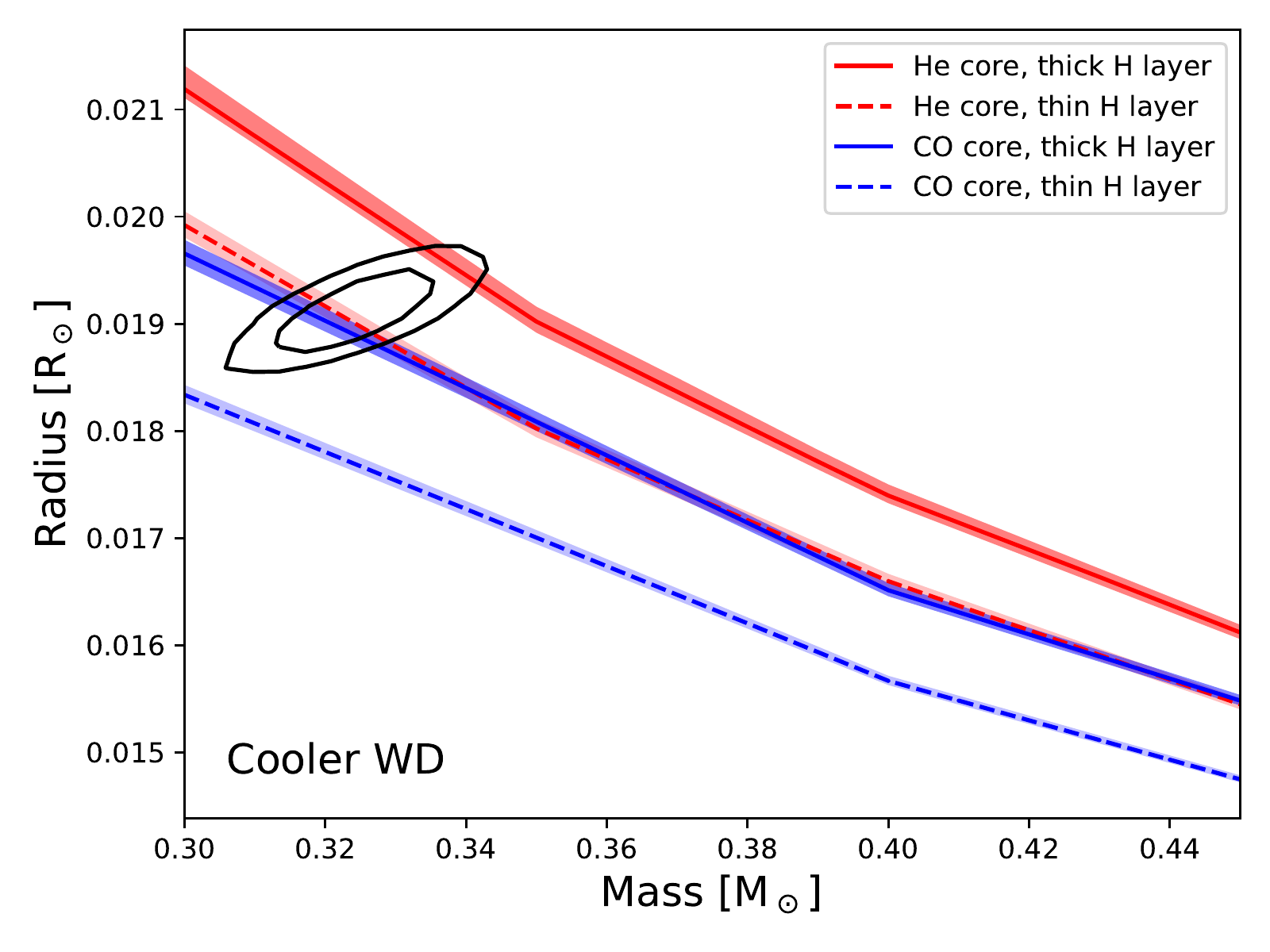}
\includegraphics[width=0.47\linewidth]{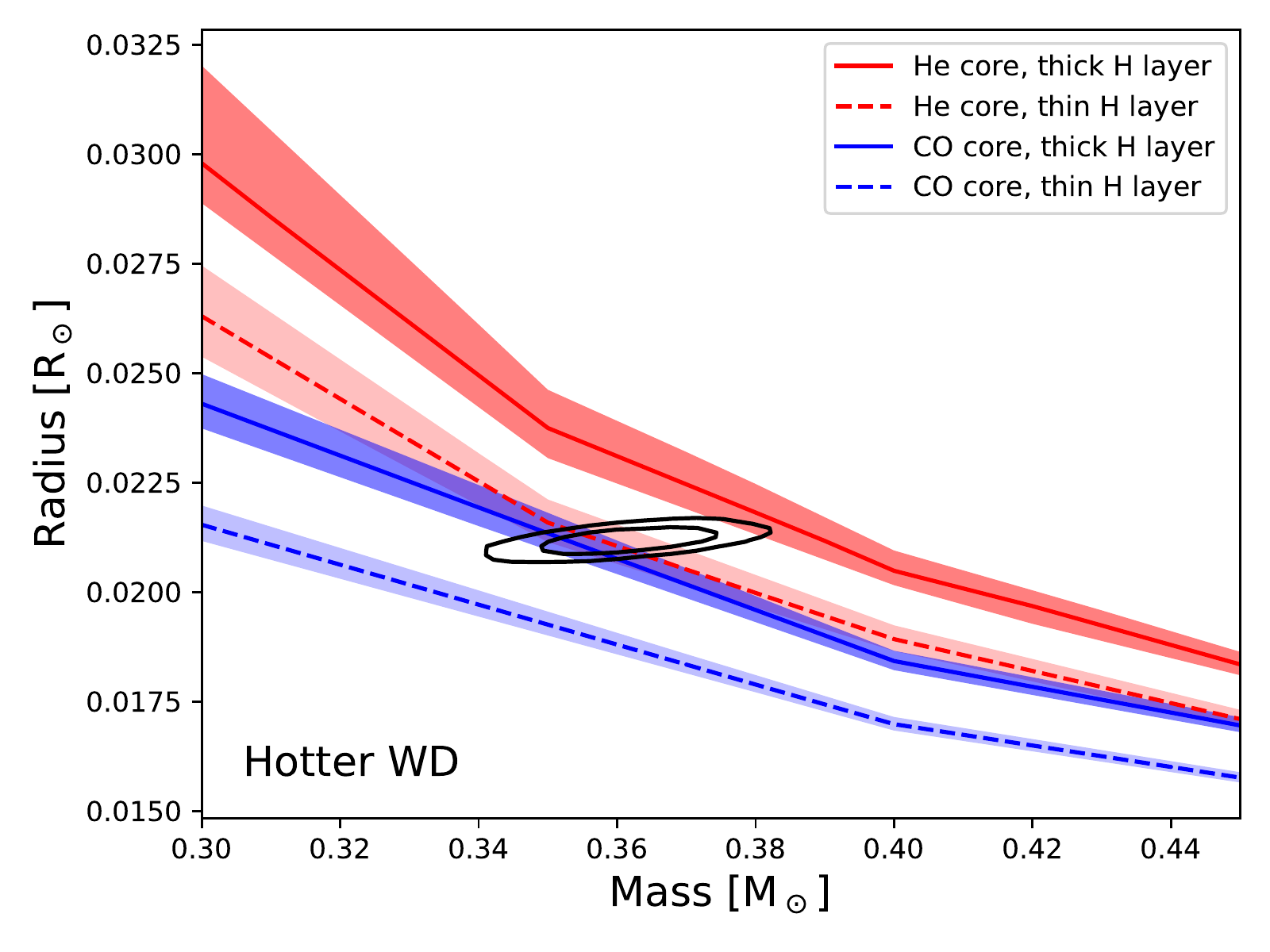}
\caption{Constraints on the masses and radii of the white dwarfs in SDSS\,J1152+0248 shown as contours (68 and 95 percentile regions). Also plotted are theoretical models for helium core white dwarfs with thick\cite{Istrate16} ($M_H/M_* = 10^{-3}$) and thin\cite{Panei07} ($M_H/M_* = 10^{-8}$) surface hydrogen layers and carbon-oxygen core white dwarfs with thick ($M_H/M_* = 10^{-4}$) and thin ($M_H/M_* = 10^{-10}$) surface hydrogen layers\cite{Fontaine01}. The theoretical models have the same temperatures as the white dwarfs in SDSS\,J1152+0248 (the shaded regions account for the uncertainty in the measured temperatures). It is also possible that these are both hybrid helium-carbon-oxygen core white dwarfs (which sit between the different models\cite{Zenati19}).}
\label{fig:mass-radius}
\end{figure}

The measured masses and radii of the white dwarfs in SDSS\,J1152+0248 are virtually model-independent (there is some small dependence upon the adopted limb darkening parameters, but this is at a level smaller than the final quoted uncertainties) and as such, comparing them to theoretical mass-radius relationships can offer insight into their internal composition\cite{Parsons17}. Figure~\ref{fig:mass-radius} shows the measured parameters of both white dwarfs in SDSS\,J1152+0248 compared to theoretical models of white dwarfs with different core compositions and surface hydrogen layer thicknesses. Given the low masses of both white dwarfs (0.36\,\msun and 0.33\,\msun) we would expect them to have helium cores\cite{Marsh95}; for example, this is the case for the similar eclipsing double white dwarf system CSS\,41177\cite{Bours14,Bours15}. However, the measured radii of both white dwarfs are too small to be consistent with pure helium core models (both white dwarfs have measured radii approximately 10\% smaller than helium core models would predict\cite{Istrate16}). On the contrary, they are more consistent with carbon-oxygen cores. While there are predictions that carbon-oxygen core white dwarfs can form at masses as low as 0.33\,\msun\cite{Prada09}, it requires substantial mass loss along the red giant branch\cite{Han00}. In order to have carbon-oxygen cores at such low masses, these white dwarfs would have needed to be initially quite massive stars ($\gtrsim 2.3$\,\msun), since they would have avoided the helium flash (as their evolution was cut short before the end of the red giant branch). This would make the system quite young as the main-sequence lifetimes of such stars is only $\sim$1\,Gyr. However, the tangential velocity of SDSS\,J1152+0248 from {\it Gaia} DR2 is more than 120\,\kms. This, combined with the large systemic velocity measured from the X-shooter data (55\,\kms) implies that the system is much older than this\cite{Tremblay14}.

Alternatively, the white dwarfs in SDSS\,J1152+0248 may have hybrid helium-carbon-oxygen cores, which have radii similar to carbon-oxygen core white dwarfs\cite{Zenati19}. It is also possible that these are helium-core white dwarfs with extremely thin surface hydrogen layers ($M_H/M_* < 10^{-8}$), although producing such a thin surface layer is difficult\cite{Istrate16}. The mass-radius measurements alone cannot distinguish between these two possibilities (helium-cores with thin surface hydrogen layers or hybrid helium-carbon-oxygen cores), however, the pulsations from the cooler white dwarf may be able to distinguish between these scenarios in the future. In particular, the mean period spacing (the period between consecutive radial overtones of a given spherical degree) will be different for alternative core compositions and is likely to be the definitive test, but this requires substantially more data than the short HiPERCAM light curve presented here. A full asteroseismic analysis, combined with the binary analysis presented in this paper, would allow us to probe the internal structure of a white dwarf that has experienced several episodes of past mass transfer within a binary system. SDSS\,J1152+0248 is therefore an ideal system for investigating how binary interactions affect the structure of white dwarfs.

\section*{Methods}

\subsection*{Observations and their reduction}

SDSS\,J1152+0248 was observed with the medium resolution echelle spectrograph X-shooter\cite{Vernet11} on the 8.2m Very Large Telescope (VLT) in Chile. Observations were performed in service mode covering one full orbit of the binary each time on 4 separate nights (March 4 2017, May 18 2017, January 17 2018 and January 20 2018). X-shooter covers a wavelength range from the UV cutoff in the blue (3000{\AA}) to the edge of the $K$-band in the red (2.5 microns) in three separate arms, the UVB (3000-5600{\AA}), VIS (5600-10100{\AA}) and NIR arms (10100-25000{\AA}). We used slit widths of 1" in the UVB arm and 0.9" in the VIS and NIR arms and binned by a factor of 2 in the spatial direction in the UVB and VIS arms, resulting in a resolution of $\sim$7500. To mitigate the effects of orbital smearing we kept exposure times short (565, 555s, 600 seconds in the UVB, VIS and NIR arms respectively). Due to the faintness of SDSS\,J1152+0248 in the NIR ($J=18.8$) these spectra have very low signal-to-noise and were therefore not used in any subsequent analysis. The data were reduced using the latest release of the X-Shooter reduction pipeline (version 3.2.0) using standard reduction steps. All spectra were placed on a barycentric wavelength scale.
 
Multi-band photometry of SDSS\,J1152+0248 was obtained with HiPERCAM\cite{Dhillon18} mounted on the 10.4m Gran Telescopio Canarias (GTC) on the 12th of January 2019. HiPERCAM is a high-speed quintuple-beam imager capable of obtaining simultaneous $u$, $g$, $r$, $i$ and $z$ band imaging at frames rates of up to 1000 frames per second with a dead time of only 0.01 seconds between exposures. HiPERCAM uses modified versions of the Sloan filters with much higher throughput, known as Super-SDSS filters. These are denoted as $u_s g_s r_s i_s z_s$ to distinguish them from the standard SDSS filters ($ugriz$). HiPERCAM allows exposure times to differ for each band, but must remain integer values of each other. Therefore, for SDSS\,J1152+0248 we used an exposure time of 2.7 seconds in the $g_s$ and $r_s$ bands, 5.4 seconds in the $u_s$ and $i_s$ bands and 8.1 seconds in the $z_s$ band, resulting in a signal-to-noise ratio of 100-200 per frame. We recorded a total of 108 minutes of data covering both a primary and secondary eclipse as well as a substantial amount of out-of-eclipse data. The data were reduced using the HiPERCAM pipeline, including fringe correction in the $z_s$ band. Differential photometry was performed using the nearby star SDSS\,J115222.49+024931.1 as a reference source. Note that this source is listed as an RR\,Lyrae star in SIMBAD, but we see no clear variability of this source during our $\sim$2 hour observations. An additional source, SDSS J115220.95+024828.4, was used as a check star to confirm the lack of variability of the comparison star. All times were converted to the barycentric dynamical timescale, corrected to the solar system barycentre, BMJD(TDB).
 
\subsection*{The radial velocity fitting method}

In order to measure the radial velocity semi-amplitudes of both white dwarfs in SDSS\,J1152+0248 we fitted the H$\alpha$ absorption line, where both components are visible (see Figure~\ref{fig:trail_fit}). The spectra were first continuum normalised by fitting a first-order polynomial to a small region either side of the H$\alpha$ line (from -2000\,\kms to -1000\,\kms and +1000\,\kms to +2000\,\kms). We then fitted the H$\alpha$ line (from -600\,\kms to +600\,\kms, hence far from the Lorentzian wings of the line) in all 62 spectra simultaneously using a combination of a first-order polynomial and three Gaussian components. We modelled the brighter white dwarf's absorption with two Gaussians, one representing the narrow non-LTE core of the line, while the other represents the broader absorption component. The non-LTE core of the H$\alpha$ line can be well approximated by a Gaussian. The fainter white dwarf was modelled with a single Gaussian component. The amplitude ($A$) and width ($\sigma$) of each Gaussian component was the same in all spectra, but their velocities ($V$) were allowed to vary from spectrum to spectrum, based on the orbital phase ($\phi$):
\begin{equation}
    V_{1,2} = \gamma_{1,2} + K_{1,2}\sin{2 \pi \phi},
\end{equation}
where $\gamma$ is the systemic velocity and $K$ is the radial velocity semi-amplitude (the subscript 1 and 2 refer to the brighter and fainter white dwarfs respectively). The two Gaussian components representing the brighter white dwarf were forced to share the same values of $\gamma_1$ and $K_1$.

The distributions of our model parameters were found using the Markov chain Monte Carlo (MCMC) method\cite{Press07} implemented using the python package {\sc emcee}\cite{Foreman13}, where the likelihood of accepting a model was based on the $\chi^2$ of the fit. A short initial MCMC chain was used to determine the approximate parameters values, which were then used as the starting values in a longer ``production'' chain, to determine the final values and their uncertainties. The production chain used 50 walkers, each with 6,000 points. The first 1,000 points were classified as ``burn-in" and were removed from the final results. The final parameters distributions are shown in Supplementary Figure~\ref{fig:trail_fit_corner}.

\subsection*{The light curve fitting method}

The HiPERCAM eclipse light curves are sensitive to the radii of the two white dwarfs, the orbital inclination and (due to the multi-band nature of HiPERCAM) also the temperatures of both white dwarfs. We generated model light curves of the system using {\sc lcurve}\cite{Copperwheat10}, which includes the effects of eclipses, Roche geometry distortion and limb- and gravity-darkening. Additionally, the code includes irradiation but this is negligible in SDSS\,J1152+0248 and so it was ignored in our modelling. {\sc lcurve} sets the surface brightnesses of the white dwarfs assuming a black body of a given temperature. However, white dwarf spectral energy distributions can depart substantially from a black body and therefore the temperatures returned by {\sc lcurve} do not accurately reflect the effective temperatures of the white dwarfs. To overcome this we computed black body temperatures for a wide range of white dwarf parameters in the $ugriz$ bands from cooling models\cite{Holberg06,Tremblay11}, in this way we could input a single effective temperature and surface gravity for each white dwarf and convert these into black body temperatures for each band before computing the light curve model.

We used the four-parameter non-linear limb darkening model\cite{Claret00}, which takes the form
\begin{equation}
\frac{I(\mu)}{I(1)} = 1 - \displaystyle\sum_{k=1}^{4}a_{k}(1-\mu^{\frac{k}{2}}),
\end{equation}
where $\mu = \cos{\phi}$ ($\phi$ is the angle between the line of sight and the emergent flux), $a_k$ are the limb darkening coefficients and $I(1)$ is the monochromatic specific intensity at the centre of the stellar disk. Coefficients for both white dwarfs were interpolated from tabulated values\cite{Gianninas13}.

The input parameters to our light curve model were the mid-time of the primary (deeper) eclipse $T_0$, the orbital period of the binary $P_\mathrm{orb}$, the binary inclination $i$, the effective temperatures of both white dwarfs $T_\mathrm{eff}$, and their masses ($M$) and radii ($R$), from which we calculate their surface gravities and thus limb-darkening coefficients and black body temperatures for each band. We also calculated the orbital separation $a$, since {\sc lcurve} takes scaled radii ($R/a$) as input parameters, rather than physical radii.

In addition to these stellar and binary parameters we also simultaneously modelled the pulsations from the cooler white dwarf with a Gaussian Process regression model implemented using the Python package {\sc celerite}\cite{Foreman17}. Gaussian process models parametrize the covariance between data points by means of a kernel function. We used a damped harmonic oscillator kernel to represent the periodic variations caused by the pulsations of the white dwarf. The kernel is defined by three parameters, the amplitude (pulse\_amp), frequency (pulse\_omega) and quality factor (pulse\_q). These parameters are fixed between the different HiPERCAM bands, but the amplitude is then re-scaled in each band based on a black body temperature (pulse\_temp).

Model parameters and their uncertainties were found using the MCMC method. However, unlike the radial velocity fitting, in this case the likelihood of accepting a model was based on a combination of the $\chi^2$ of the fit to the light curve data and an additional prior probability to ensure that the masses were consistent with the measured radial velocities. Moreover, the previous eclipse timing data\cite{Hallakoun16} were included to improve the accuracy and precision of the ephemeris. As before, initial chains were used to determine the starting parameters for a full production chain. This final chain contained 50,000 points, of which the first 5,000 were excluded as burn-in. The final parameters distributions are shown in Supplementary Figure~\ref{fig:lcurve_fit_corner}. 

While our results are generally in agreement with the original study of this system\cite{Hallakoun16}, the temperatures for both white dwarfs are significantly lower, as is the mass of the cooler component. All of these discrepancies are due to an overestimation of the contribution from the cooler component in the original study. In the original study the cooler component was found to contribute 31\% of the total $g$ band flux. Our higher quality data shows that its contribution is 22.4\% in the $g$ band. Their higher contribution was then folded into the analysis of the spectral energy distribution in the original study, resulting in an overestimation of the temperatures of both white dwarfs. This was compounded by the lack of {\it Gaia} data to constrain the distance and the fact that they only had eclipse data in a single photometric band. In the original study the mass of the cooler white dwarf was then determined using a mass-radius relationship for a fixed temperature. However, as this temperature was too high it caused an overestimation of the mass (since at a fixed temperature higher mass white dwarfs have smaller radii). Lower mass white dwarfs were excluded since their radii would have been too large at these hotter temperatures. However, the methodology used in that study is still sound; had they used the temperatures that we measured then their analysis would have yielded a similar mass for the cooler white dwarf (since a 10,500\,K 0.33\,{\msun} white dwarf has a similar radius to a 14,500\,K 0.44\,{\msun} white dwarf).

\subsection*{The spectral energy distribution fitting method}

In addition to constraining the temperatures of both white dwarfs via the multi-band eclipse light curves, we also performed a fit to the spectral energy distribution of SDSS\,J1152+0248 in order to measure independently the temperatures via this method. We fitted the GALEX, SDSS and UKIDSS photometry with a combination of two model DA white dwarf spectra\cite{Koester10}. The input parameters for the fit were the temperatures and surface gravities of both white dwarfs and the parallax and reddening $E(B-V)$, of the system. The model spectra for each white dwarf were scaled to the distance implied from the parallax (by simple parallax inversion and using a mass-radius relationship\cite{Fontaine01}), reddened, and combined. 
Model parameters and their uncertainties were again found using the MCMC method. We included an additional prior probability based on the {\it Gaia} parallax\cite{Gaia18} and set an upper limit on the reddening of $E(B-V)<0.106$ based on reddening maps\cite{Schlafly11}. Our prior also ensured that the masses were consistent with the spectroscopic values and that the relative contribution of each white dwarf in the SDSS $ugriz$ bands were consistent with the measurements from the eclipse light curves. We again used initial chains to determine the starting parameters for the final production chains. The final chain contained 10,000 points and the first 2,000 were excluded as burn-in. The final parameters distributions are shown in Supplementary Figure~\ref{fig:sed_fit_corner}.

\subsection*{Pulsation analysis}

We explored possible oscillation frequencies of the binary-light-curve-model-subtracted residuals using the {\sc Period04} software package\cite{Lenz05}. For each of the five bandpasses we determined what signals in the periodogram were significant by determining a 1\% false-alarm probability. To do so we kept the time sampling of the light curve the same but randomly shuffled the flux values $10{,}000$ times, taking the top 1\% of the cumulative distribution of highest peaks from each synthetic, shuffled light curve as the 1\% false-alarm probability\cite{Hermes17}. The 1\% false-alarm probabilities are shown in Figure~\ref{fig:pulsations} ($u_s$: 3.28\,ppt (parts per thousand); $g_s$: 1.41\,ppt; $r_s$: 1.97\,ppt; $i_s$: 3.64\,ppt; $z_s$: 9.20\,ppt), and we consider any peaks in the periodogram above these values to be significant. One significant signal is detected in $u_s$ at $1300\pm18$\,s ($42.9\pm5.4$\,ppt amplitude). Three significant signals are detected in $g_s$ at $1314.0\pm5.9$\,s ($33.0\pm1.3$\,ppt amplitude), $1059\pm13$\,s ($9.8\pm1.3$\,ppt amplitude), and $582.9\pm4.3$\,s ($8.9\pm1.3$\,ppt amplitude). Two significant signals are present in $r_s$ at $1319.4\pm9.5$\,s ($25.3\pm1.6$\,ppt amplitude) and $1050\pm17$\,s ($9.1\pm1.6$\,ppt amplitude). One significant signal is present in $i_s$ at $1320\pm19$\,s ($22.7\pm2.9$\,ppt amplitude). There are no significant signals in $z_s$. The reported pulsation amplitudes have been adjusted from those measured and shown in Figure~\ref{fig:pulsations} to account for the flux contribution of the pulsating secondary white dwarf in each bandpass ($u_s$: $12.3\pm0.5$\%; $g_s$: $18.4\pm0.2$\%; $r_s$: $22.4\pm0.2$\%; $i_s$: $24.8\pm0.2$\%; $z_s$: $26.4\pm0.3$\%). We have adopted the pulsation periods determined from the $g_s$ in the manuscript, since they have the highest signal-to-noise.

\section*{Data availability}

The raw and reduced X-shooter data presented in this paper are available from the ESO archive: \url{http://archive.eso.org/cms.html}. Raw and reduced HiPERCAM data are available from the GTC archive: \url{http://gtc.sdc.cab.inta-csic.es/gtc/index.jsp}. These data can also be obtained from the corresponding author upon reasonable request.

\section*{Code availability}

The X-Shooter reduction pipeline (version 3.2.0) is available at \url{https://www.eso.org/sci/software/pipelines/xshooter/} and the HiPERCAM pipeline at \url{https://github.com/HiPERCAM/hipercam}. The light-curve fitting method is available at \url{https://github.com/trmrsh/cpp-lcurve}. The codes used to generate the plots presented in this paper are available from the corresponding author upon reasonable request.

\bibliography{sdss1152}

Correspondence and requests for materials should be addressed to Dr. S. G. Parsons (s.g.parsons@sheffield.ac.uk)

\section*{Acknowledgements}

S.G.P. acknowledges the support of a Science and Technology Facilities Council (STFC) Ernest Rutherford Fellowship. HiPERCAM and V.S.D. are funded by the European Research Council under the European Union’s Seventh Framework Programme (FP/2007-2013) under ERC-2013-ADG Grant Agreement no. 340040 (HiPERCAM). Partial support for this work was provided by NASA K2 Cycle 6 Grant 80NSSC19K0162. A.G.I acknowledges support from the Netherlands Organisation for Scientific Research (NWO).

Based on observations made with the Gran Telescopio Canarias (programme ID\,GTC59-18B), installed in the Spanish Observatorio del Roque de los Muchachos of the Instituto de Astrof\'isica de Canarias, in the island of La Palma. Based on observations made with ESO Telescopes at the La Silla Paranal Observatory under programme ID\,097.D-0786.

\section*{Author contributions statement}

All authors have contributed to the work presented in this paper. S.G.P. reduced all the spectroscopic and photometric data and carried out the SED fitting. S.G.P. and V.S.D. performed the GTC observations. A.J.B. performed the radial velocity and light curve fitting. S.P.L. wrote the Python code that implemented Gaussian Processes in the light curve fitting. V.S.D., S.P.L., T.R.M., S.G.P., E.B. M.J.D., M.J.G. and D.I.S. all contributed to the development and support of HiPERCAM. J.J.H. analysed the pulsations in the HiPERCAM light curves. A.G.I. investigated the internal structure of the white dwarfs and the evolution of the binary. All authors reviewed the manuscript.

\section*{Additional information}

\textbf{Competing Interests}. The authors declare that they have no competing financial interests.

\renewcommand{\figurename}{Supplementary Figure}

\renewcommand\thefigure{\arabic{figure}}    

\setcounter{figure}{0}    

\begin{figure}[ht]
\centering
\includegraphics[width=0.95\linewidth]{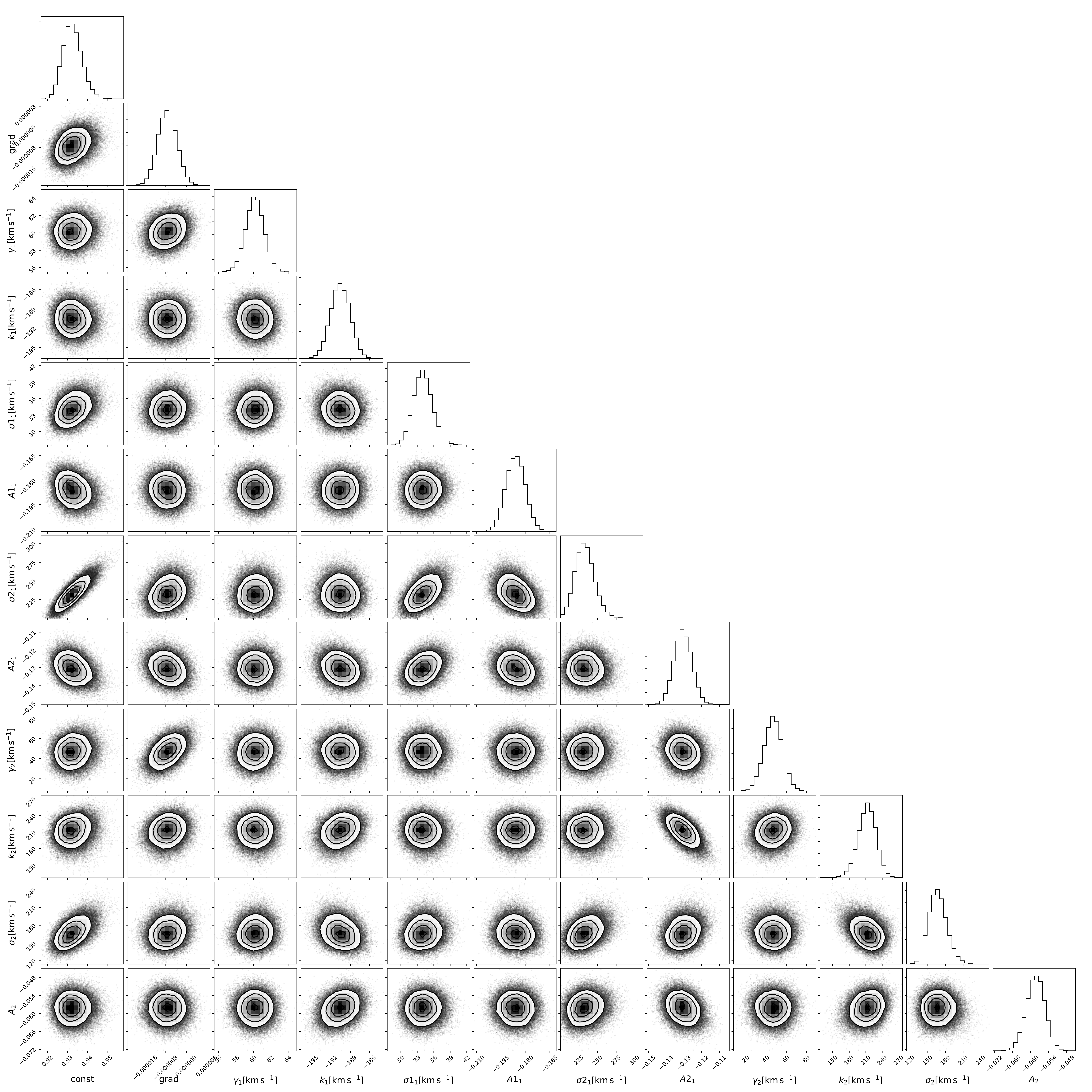}
\caption{Posterior probability distributions for model parameters obtained through fitting the H$\alpha$ absorption lines of both white dwarfs. Grey-scales and contours represent the joint probability distributions for each pair of parameters , while the histograms show the marginalised probability distributions for each parameter.} 
\label{fig:trail_fit_corner}
\end{figure}

\begin{figure}[ht]
\centering
\includegraphics[width=0.95\linewidth]{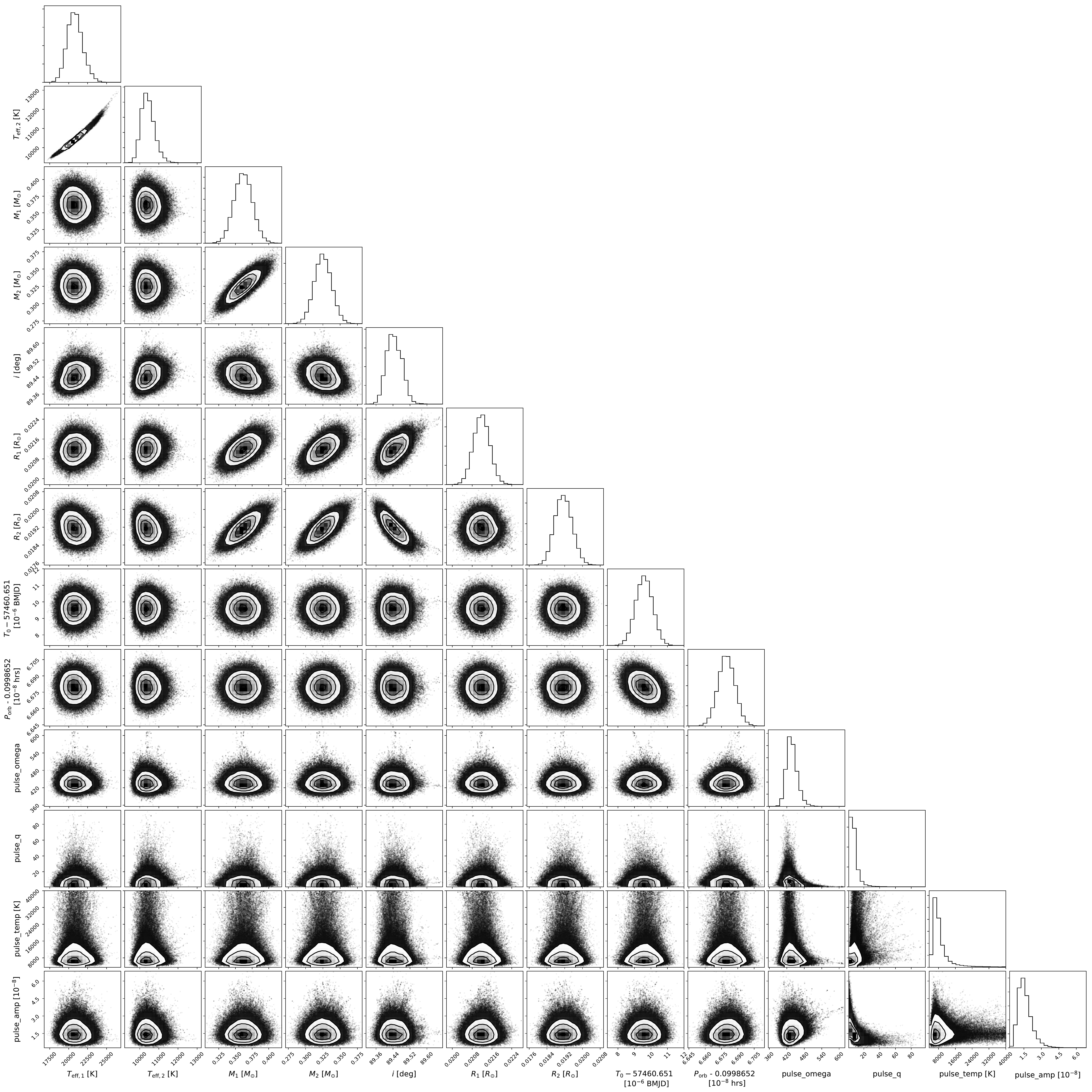}
\caption{Posterior probability distributions for model parameters obtained through fitting the multi-band HiPERCAM light curves.} 
\label{fig:lcurve_fit_corner}
\end{figure}

\begin{figure}[ht]
\centering
\includegraphics[width=0.95\linewidth]{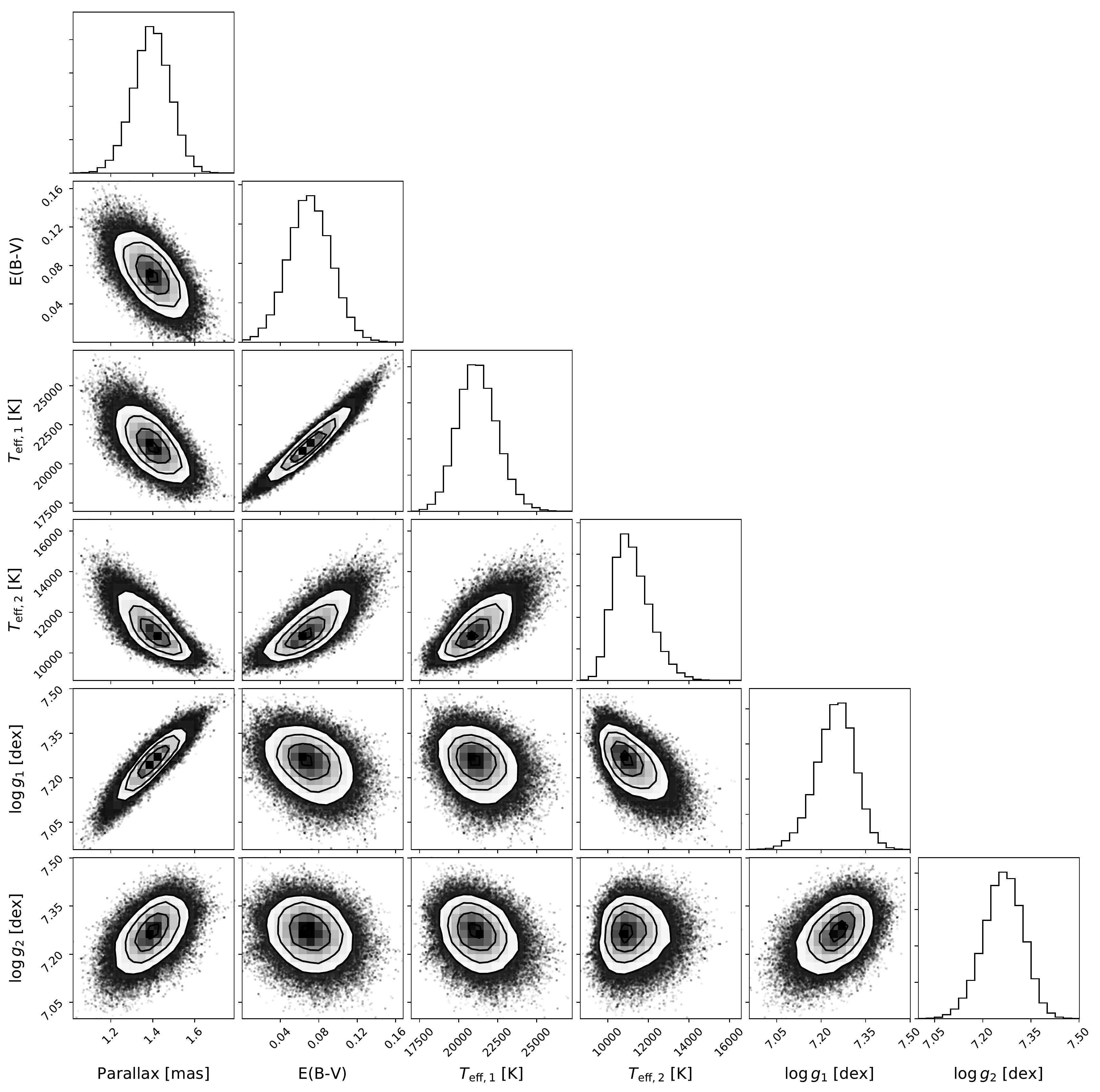}
\caption{Posterior probability distributions for model parameters obtained from the spectral energy distribution fit.} 
\label{fig:sed_fit_corner}
\end{figure}

\end{document}